\documentclass[preprint,5p]{elsarticle}

\usepackage[english]{babel}

\usepackage[utf8]{inputenc}
\usepackage{natbib}
\usepackage{graphicx}
\usepackage{amsmath}
\usepackage{amssymb}
\usepackage{caption}
\usepackage{siunitx}
\DeclareSIUnit\atomicmassunit{u}
\usepackage{subcaption}
\usepackage{booktabs}
\usepackage{multirow}
\usepackage{hyperref}

\newcommand{\replyTo}[1]{\textit{** Comment #1 ** }}
\renewcommand{\replyTo}[1]{}

\newcommand{\replyToNew}[1]{\textit{** Comment #1 ** }}
\renewcommand{\replyToNew}[1]{}

\journal{Life Sciences in Space Research}

\begin{document}
\begin{frontmatter}
\title{Hybrid Active-Passive Galactic Cosmic Ray Simulator: \\experimental implementation and microdosimetric characterization}

\author[1]{E.~Pierobon}
\ead{E.Pierobon@gsi.de}
\author[1,2]{L.~Lunati}
\ead{L.Lunati@gsi.de}
\author[1]{T.~Wagner}
\ead{T.Wagner@gsi.de}
\author[1,2,4]{M.~Durante}
\ead{M.Durante@gsi.de}
\author[1]{C.~Schuy\texorpdfstring{\corref{cor1}}{}}
\ead{C.Schuy@gsi.de}

\cortext[cor1]{Corresponding author}

\affiliation[1]{organization={Biophysics Department, GSI Helmholtzzentrum für Schwerionenforschung},
    addressline={Planckstraße 1},
    postcode={64291},
    city={Darmstadt},
    country={Germany}}
\affiliation[2]{organization={Institute of Condensed Matter Physics, Technische Universität Darmstadt},
    addressline={Hochschulstr. 6},
    postcode={64289},
    city={Darmstadt},
    country={Germany}}
\affiliation[4]{organization={Department of Physics ``Ettore Pancini'', University Federico II}, 
    addressline={Via Cintia, 21 - Building 6},
    postcode={80126},
    city={Naples},
    country={Italy}}

\begin{abstract}
Space radiation is one of the major obstacles to space exploration.
If not mitigated, radiation can interact both with biological and electronic systems, inducing damage and posing significant risk to space missions. 
Countermeasures can only be studied effectively with ground-based accelerators that act as a proxy for space radiation. \\
Following an in-silico design and optimization process, we have developed a galactic cosmic ray (GCR) simulator using a hybrid \textit{active-passive} methodology. In this approach, the primary beam energy is \textit{actively} switched and the beam interacts with specifically designed \textit{passive} modulators.\\
In this paper, we present the implementation of such a GCR simulator and its experimental microdosimetric characterization.
Measuring the GCR field is of paramount importance, both before providing it to the user as a validated radiation field and for achieving the best possible radiation description.
The issue is addressed in this paper by using a tissue equivalent proportional counter to measure radiation quality and by comparing experimental measurements with Monte Carlo simulations.
In conclusion, we will demonstrate the GCR simulator's capability to reproduce a GCR field.
\end{abstract}

\begin{keyword}
Space Radiation \sep Galactic Cosmic Ray simulator \sep Microdosimetry \sep GCR \sep Monte Carlo
\end{keyword}

\end{frontmatter}
\section{Introduction}

Space radiation is acknowledged as one of the main hindrances to manned exploration of the Solar system and poses one of the major scientific challenges for planned long-term missions to Moon and Mars \cite{Chancellor2014,Durante2014,Durante2011,Patel2020,Bahadori2024,Fogtman2023}.\\
The highly energetic, heavy ions of the galactic cosmic rays (GCRs) have no natural counterpart on Earth and can only be studied in-silico, directly in space or with the aid of high-energy heavy ion accelerators \cite{Slaba2016,Durante2005}. Systematic studies of the effects of energetic heavy ions on, e.g., biological systems are typically performed at particle accelerators with one or a small subset of energies and ion species as a proxy for the complex space radiation environment \cite{Giraudo2018,Luoni2022,Kodaira2024}.\\
To facilitate scientific studies of these effects, the National Aeronautics and Space Administration (NASA) developed an irradiation system at the Brookhaven National Laboratory able to simulate a GCR spectrum \cite{Simonsen2020, norbury2016galactic}. In Europe, the GSI Helmholtzzentrum für Schwerionenforschung (GSI), in collaboration with the European Space Agency (ESA), has developed a hybrid active-passive approach to improve the ground-based simulation of space radiation \cite{Schuy2020}. 
\replyTo{17}Both facilities offer a unique platform for studying the effects of GCR on ground, featuring distinctive features that are specific to their design and implementation choices.
The design and optimization of the hybrid active-passive GCR simulator is described in detail in \cite{lunati2025mc} and only briefly summarized in the following. To reproduce a GCR field, the approach employs the sequential irradiation of six configurations, with each configuration varying in primary beam ($^{56}\text{Fe}$) energy and beam modulator: 
\begin{enumerate}
    \item Complex modulator at \qty{1}{\GeV \per \atomicmassunit},
    \item Complex modulator at \qty{0.7}{\GeV \per \atomicmassunit},
    \item Complex modulator at \qty{0.35}{\GeV \per \atomicmassunit},
    \item Slab modulator composed of \qty{80}{\mm} of steel-304L in combination with the FRAgment kiNetiC energy Optimizer (FRANCO) at \qty{1}{\GeV \per \atomicmassunit} primary beam energy,
    \item Slab modulator composed of \qty{50}{\mm} of steel-304L in combination with the FRANCO at \qty{1}{\GeV \per \atomicmassunit} primary beam energy,
    \item Two slab modulators: composed of \qty{50}{\mm} of steel-304L and \qty{100}{\mm} of polyethylene (PE) at \qty{0.35}{\GeV \per \atomicmassunit} primary beam energy.
\end{enumerate}
\replyTo{17}Following the optimization process detailed in \cite{lunati2025mc}, a weight is assigned to each configuration, thus establishing a link between the number of primary particles required in each configuration.
This sequential irradiation is then expected to reproduce the GCR field after \qty{10}{\g \per \cm \squared} of aluminum shielding, at \qty{1}{\astronomicalunit} and in solar minimum conditions (2010 solar minimum). \\
\replyTo{17}Microdosimetry is an invaluable tool for characterizing mixed radiation fields. 
Due to its defining characteristic of measuring the energy deposition at the micrometric scale, embedding the stochastic nature of energy deposition, microdosimetry provides an accurate description of the quality of the radiation field\cite{Griffiths1985,Missiaggia2024}.
Tissue equivalent proportional counters (TEPCs) can be considered the reference detectors for microdosimetry. TEPCs are not new to space applications as their response to heavy ions has been studied in detail \cite{Rademacher1998,Gersey2002,Guetersloh2004,Miller2016} and they have been deployed to characterize the space radiation environment in multiple missions \cite{Badhwar1995,Zhou2007,Doke2001}. 
\replyTo{17}Furthermore, their tissue equivalence, one of their defining features alongside appropriate modeling, allows physical descriptions to be linked more directly to biological effects. \\
This manuscript deploys a TEPC microdosimeter to characterize and validate the complex radiation field specific to each of the six experimental configurations.
A comparison is made between experimentally-derived and in silico data, examining the $f(y)$ distributions.
Following the optimization process delineated in \cite{lunati2025mc}, the GCR simulator spectrum is subsequently determined employing the designated weights for each configuration.
A comparison is made between the resulting GCR simulator spectrum and state-of-the-art Monte Carlo data. 
Given that the optimization process employed in the design of the GCR simulator is exclusively based on physical-derived quantities, without directly accounting for biological parameters, it is anticipated that, if the adopted methodology is valid, biological parameters will align accordingly.
Consequently, the quality factor $Q$ is calculated from microdosimetric spectra as it can be used to gain valuable insight into the effects of radiation on biological systems \cite{KELLERER1985,Kellerer1978,Rossi1996}.
$Q$ typically represents an average of the maximum relative biological effectiveness (RBE) values associated with stochastic effects, such as cancer incidence \cite{Valentin2003,Kyriakou2021}.
\replyTo{17}$Q$ is then compared against the designed quality factor obtained from the optimization process in \cite{lunati2025mc}.
\replyTo{8+17}Finally, microdosimetric measurements conducted in low Earth orbit (LEO) are used as a proxy for deep space radiation environments. Although the conditions in LEO differ from those in deep space, these measurements are indispensable for advancing space exploration research \cite{mcphee2025iss4mars, NARICI2018990}.
The microdosimetric spectrum obtained by the GCR simulator is then compared with an independently acquired microdosimetric spectrum from Space Shuttle Mission STS-102 to further validate and demonstrate the efficacy of the developed GCR simulation.\\
\replyTo{17}The GCR simulator at GSI is poised to make a significant impact on the field of space-related studies. Its unique methodology sets it apart from other simulators and will provide a groundbreaking platform for advancing the frontiers of space exploration.

\section{Methods}
\subsection{Hybrid active-passive galactic cosmic ray simulator implementation}
\begin{figure*}[htb]
    \centering
    \includegraphics[width=\textwidth]{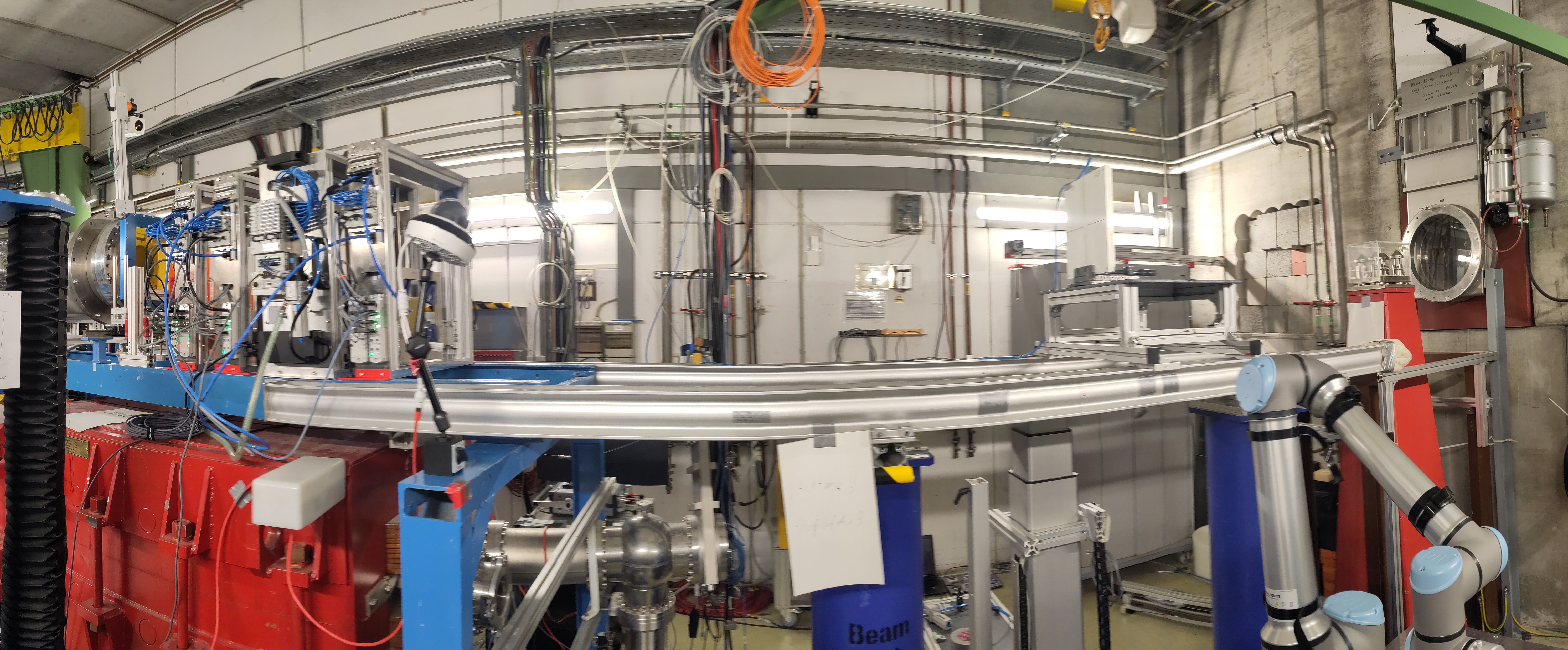}
    \caption{Photo of the experimental setup implemented in Cave A.}
    \label{fig:methods:picture_of_the_setup}
\end{figure*}

\begin{figure*}[htb]
    \centering
    \includegraphics[width=\textwidth]{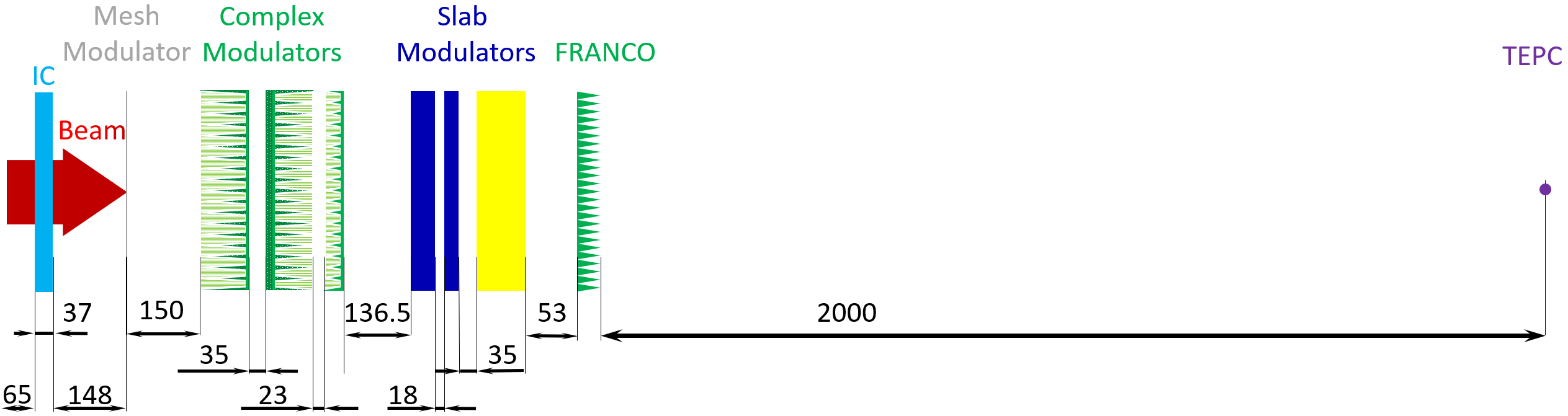}
    \caption{Schematic of the experimental setup of the GCR simulator implemented in Cave A for the TEPC measurement. 
    The picture, for illustrative purposes, shows all the elements in the beam path. Once a GCR simulator configuration has been selected, the elements are removed/inserted into the beamline via remote controllable modulator exchangers accordingly.
    Item colors are assigned based on their respective materials. Vertical axis is not to scale, all units are in \unit{\mm}.}
    \label{fig:methods:scheme_exp_setup}
\end{figure*}

The hybrid active-passive galactic cosmic ray simulator (hereafter simply the GCR simulator) setup has been implemented in the beamline of GSI's Cave A  and it is shown in \autoref{fig:methods:picture_of_the_setup}, whereas all distances and individual components are presented in \autoref{fig:methods:scheme_exp_setup} and briefly described in the following.


\subsubsection{Beam application and monitoring}
The primary beam provided by the accelerator is monitored by a large-area parallel-plate ionization chamber (IC). 
Ions passing through the IC generate a charge, which is converted into pulses with a charge to frequency converter \cite{Luoni2020}. The relationship between the IC pulses and the number of primary particles is established via a cross calibration with a plastic scintillator. The IC pulses are then used to monitor the irradiation and control the magnet scanning for homogeneously covering the passive beam modulators. 
Beam was delivered via the raster scanning system, available in cave A, to span a uniform square area during irradiation.

\subsubsection{Modulators}

The GCR simulator design heavily relies on different forms of passive beam modulators, namely a \textit{mesh modulator}, \textit{complex modulators}, where the geometrical shape is optimized, and \textit{slab modulators}, where only the thickness is optimized. Each type of modulator fulfills a well-defined role described in detail in \cite{lunati2025mc}. To allow automated switching between different configurations, all modulators are mounted on remote controllable modulator exchangers.

\paragraph{Mesh modulator}
Regardless of the irradiated configuration, the primary beam always interacts with a 32-layer steel-304L mesh modulator \cite{Tanaka2022}. Each layer is composed of wires with a diameter of \qty{60}{\micro \m} and a pitch of \qty{75}{\micro \m}. The layers are mounted in a 3D-printed frame with random orientations to form a pseudo-random structure, while ensuring a nearly uniform effective thickness.

\paragraph{Complex and fragments kinetic energy optimizer modulators}
\label{ref:introduction:complex_mod}
Complex modulators are produced in-house on a 3DSystems ProJet MJP 2500 Plus 3D-printer. The printing material used is \textit{VisiJet M2S-HT250} with \textit{VisiJet M2 SUP} as support material. The printing material has a density of \qty{1.1819}{\g \per \cm \cubed }. Following the printing process the modulators are post-processed for support material removal.

\paragraph{Slab modulators}
Optimized slab modulators can be directly machined out of suitable materials. Steel-304L with a nominal density of \qty{8}{\g \per \cm \cubed} was used as material for the steel slab modulators while polyethylene (PE) with a nominal density of \qty{0.94}{\g \per \cm \cubed} was used for the \qty{100}{\mm} PE slab modulator. 

\subsubsection{Target handling}
All targets to be irradiated using the GCR simulator are positioned by a robotic arm (Universal Robots UR10e), 
equipped with a manual precision tool changer (Grip GmbH). 
The UR10e offers a maximum payload of \qty{12.5}{\kg} and an extended reach of up to \qty{1.3}{\m}.
The robotic arm’s extension is required to position targets before and after irradiation in an automated fashion at a safe distance from the beamline, thereby reducing the impact of stray radiation. 
Specifically designed holders are used to mount targets to the robotic arm while biological samples are typically manipulated with soft grippers (SoftGripping GmbH).

\subsection{Tissue equivalent proportional counter detector}
To characterize and validate the radiation field a commercially available tissue equivalent proportional counter (Far West Technology LET-1/2) was deployed. 
It was filled with propane gas at density of \qty{1.08e-4}{\g \per \cm \cubed} simulating a tissue equivalent spherical active volume with a diameter of \qty{2}{\micro \m} and aligned via the robotic arm in the target area \qty{2}{\m} behind the last beamline element, centered to the beam direction.
The TEPC was biased to \qty{700}{\V} via a high voltage power supply (ISEG SHR 4060R) while the output signal was processed by a preamplifier (model 2006 from Camberra).
The preamplified signal was split and sent to three different shaping amplifiers with gain settings of approximately $10$ (model 671 from Ortec), $100$ (model 7234E from ISEG), and $1000$ (model 757A from Ortec).
Shaping time was selected as the best compromise according to the environmental noise and has been fixed to a value of \qty{1}{\micro \s} for the ${\sim} 10$- and ${\sim} 100$-gain settings while an optimal value of \qty{1.5}{\micro \s} for the ${\sim} 1000$-gain setting was found.
Signals were digitized by two, two-channel multichannel analyzers (ORTC Model 972) at 14 bits, connected to a PC operated remotely from the cave control room.
Electronic noise was affecting only the low channels of the ${\sim} 1000$-gain amplifier and was removed by setting a minimum channel threshold in the data acquisition software (ORTEC Maestro). 
The aforementioned acquisition chain produces three spectra, one for each level of gain, which are joined and processed to create a single microdosimetric spectrum.\\
To calibrate the lineal energy of the detector, microdosimetric spectra with pure monoenergetic iron beam were acquired at \qty{1}{\GeV \per \atomicmassunit}, \qty{0.7}{\GeV \per \atomicmassunit} and \qty{0.35}{\GeV \per \atomicmassunit}.
These were directly compared to the MC simulation applying a similar procedure as described in \cite{Missiaggia2020, Pierobon2023}.

Microdosimetric spectra and quantities were calculated according to the definition in \cite{Griffiths1985}.
Similarly, following the definition proposed by ICRU 40 \cite{ICRU40} revisited by Kellerer and Hahn \cite{Kellerer1988}, the quality factor $Q_y(y)$ was calculated from the $d(y)$ microdosimetric spectrum.
An additional definition provided by ICRP 60 \cite{ICRP60} of quality factor $Q_L(L)$ can be calculated assuming the lineal energy $y$ as a descriptor of the Linear Energy Transfer (LET).
Both definitions are reported in \autoref{eq:methods:quality_KH} and \autoref{eq:methods:quality_ICRP} respectively.
\begin{subequations}
\label{eq:methods:quality_definition}
\begin{align}
    Q_y(y) & \equiv 0.3 y  \left[ 1 + \left(\frac{y}{137}\right)^5 \right]^{-0.4}, \label{eq:methods:quality_KH} \\
    Q_L(L) & \equiv \begin{cases}
        1, & L \leq \qty[per-mode = fraction]{10}{\keV \per \micro \m} \\ 
        0.32L - 2.2, & \qty[per-mode = fraction]{10}{\keV \per \micro \m}  < L \leq \qty[per-mode = fraction]{100}{\keV \per \micro \m}  \\ 
        \frac{300}{\sqrt{L}}, & L > \qty[per-mode = fraction]{100}{\keV \per \micro \m} \label{eq:methods:quality_ICRP}
    \end{cases} \, .
\end{align}
\end{subequations}

An individual experimental measurement was acquired for each of the six GCR simulator configurations. 
Regarding the three complex modulator configurations, a scan field size of \qtyproduct{100 x 100}{\mm} was used, while, for the slab modulators, a smaller field of \qtyproduct{98 x 98}{\mm} was used.
The reason for this was to compensate for the divergence of the raster scanning system in the slab modulator configurations, which were positioned at a greater distance than the complex modulator.
The beam scanning area was verified with the aid of gafchromic films.
Before any measurement, the particles rate was adjusted to a sustainable and conservative value for the detector of roughly $\lesssim 100$ events per second.
\replyTo{16}At least $2 \times 10^6$ primary particles were delivered for each configuration. 

\subsection{Monte Carlo simulations}
\label{ssec:Monte Carlo simulations}
\begin{figure*}[htb]
    \centering
    \includegraphics[width=\textwidth]{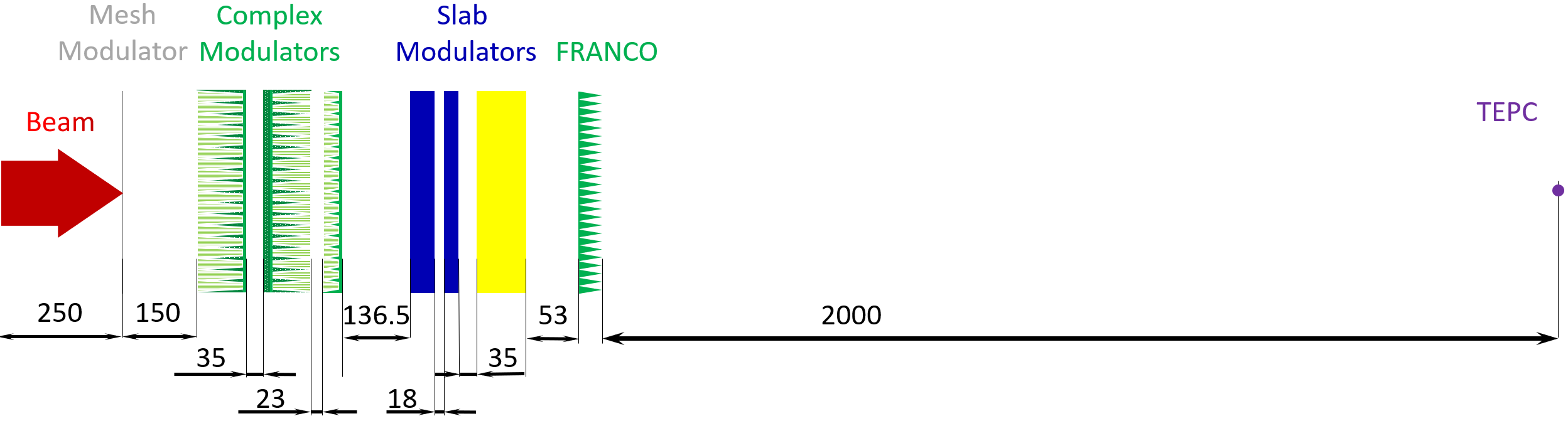}
    \caption{Scheme of the setup implemented in the Monte Carlo simulation.
    The picture, for illustrative purposes, shows all the elements in the beam path. 
    Upon selecting a GCR simulator configuration, only the required elements are initialized accordingly.
    Item colors are assigned to their respective materials. 
    Vertical axis is not to scale, all units are in $\si{mm}$.}
    \label{fig:methods:MC}
\end{figure*}

To validate the GCR simulator and to compare the experimental data against simulations results, the GCR simulator beamline was implemented in Geant4 Monte Carlo toolkit version 11.2.1 \cite{Agostinelli2003, Allison2006, Allison2016}.
Hadronic interaction was driven by the physics list \texttt{QBBC} which is recommended for space-related applications and electromagnetic interaction was modeled by \texttt{option3} \replyTo{3}(EMY). 
A schematic drawing of the MC setup can be found in \autoref{fig:methods:MC}.
Distances and dimensions are matching the ones measured in the experimental setup of \autoref{fig:methods:picture_of_the_setup} and in \autoref{fig:methods:scheme_exp_setup}.
The materials definition in the simulations were implemented based on the National Institute of Standard and Technology (NIST) database if available, otherwise, following the manufacturer's specifications.
Each modulator has a transverse area of \qtyproduct{100 x 100}{\mm}, is centered on the beamline, and is oriented orthogonally to the beamline.
Furthermore, the simulation models the TEPC LET-1/2 detector at \qty{2}{\m} from the last element of the beamline. 
The latter was implemented following a similar approach as in \cite{Missiaggia2021} with further details of the detector geometry and composition available in \ref{sec:MC_implementation_of_the_TEPC}.

Primary beam was modeled by a monoenergetic source with a specific primary energy depending on the configuration. To mimic the raster scanning system, the beam was generated from a uniform area of \qtyproduct{80 x 80}{\mm} originating at world edge. 
\replyTo{7}To reduce the computational complexity and gain efficiency, the simulated field size was reduced compared to the experimental field size.
\replyTo{20}No lateral beam divergence was added to the simulations because the mesh modulator interacts the primary beam inducing sufficient lateral scattering and energy straggling \cite{lunati2025mc}.
\replyTo{16}A minimum number of $1 \times 10^7$ events were generated for each configuration.

\section{Results}
\subsection{Microdosimetric spectra}
\begin{figure*}[ht]
    \centering
    \begin{subfigure}[t]{0.32\textwidth}
        \centering
        \includegraphics[width=\textwidth]{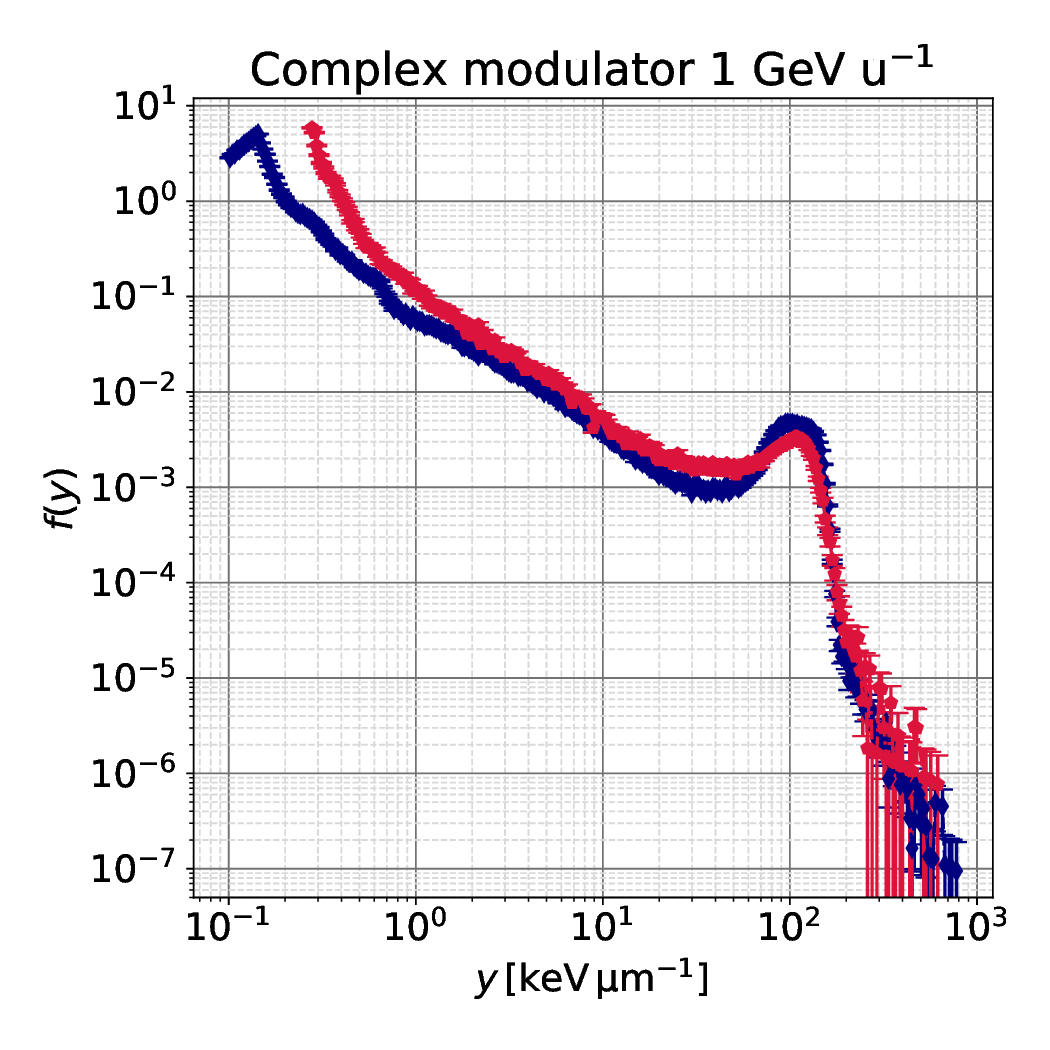}
        \caption{}
        \label{fig:results:1GeV_mod}
    \end{subfigure}
    \begin{subfigure}[t]{0.32\textwidth}
        \centering
        \includegraphics[width=\textwidth]{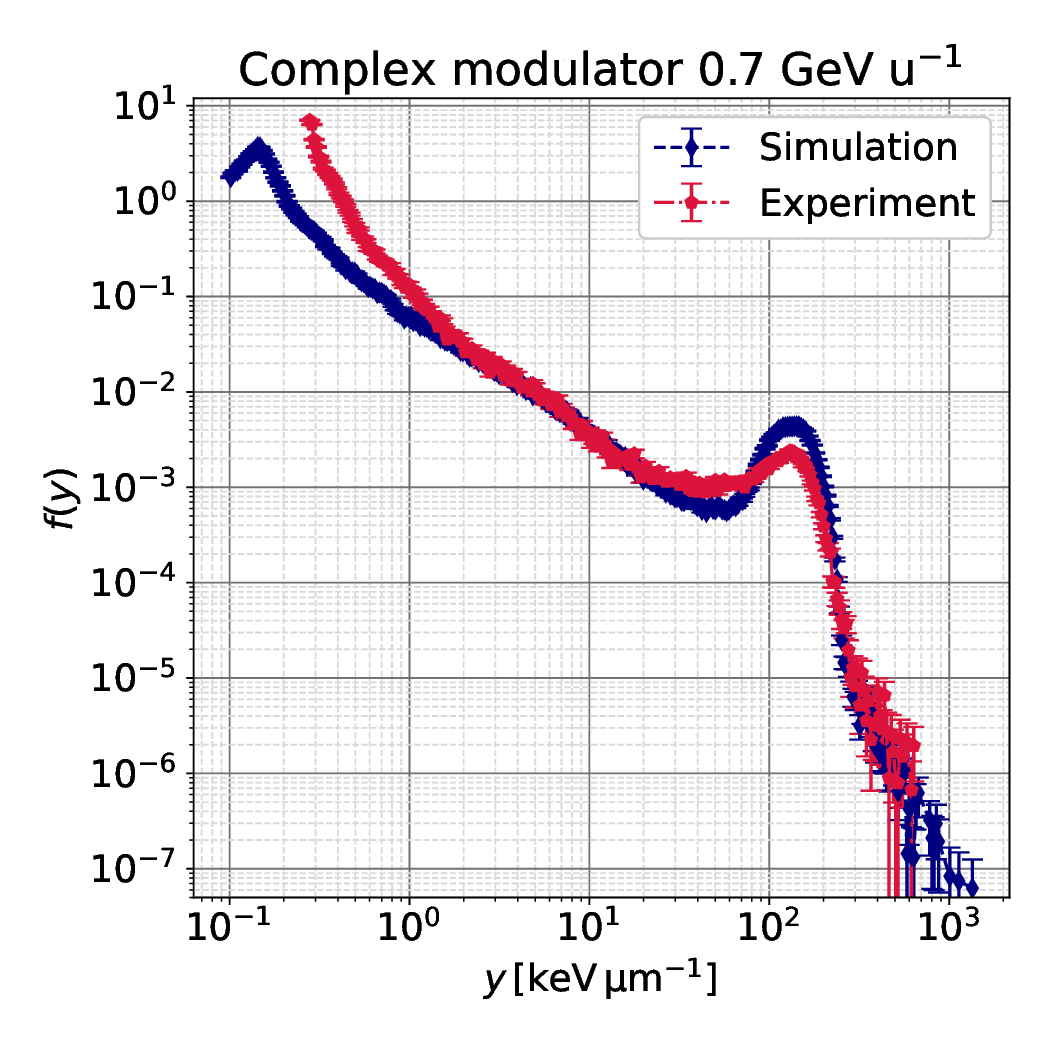}
        \caption{}
        \label{fig:results:0.7GeV_mod}
    \end{subfigure}
    \begin{subfigure}[t]{0.32\textwidth}
        \centering
        \includegraphics[width=\textwidth]{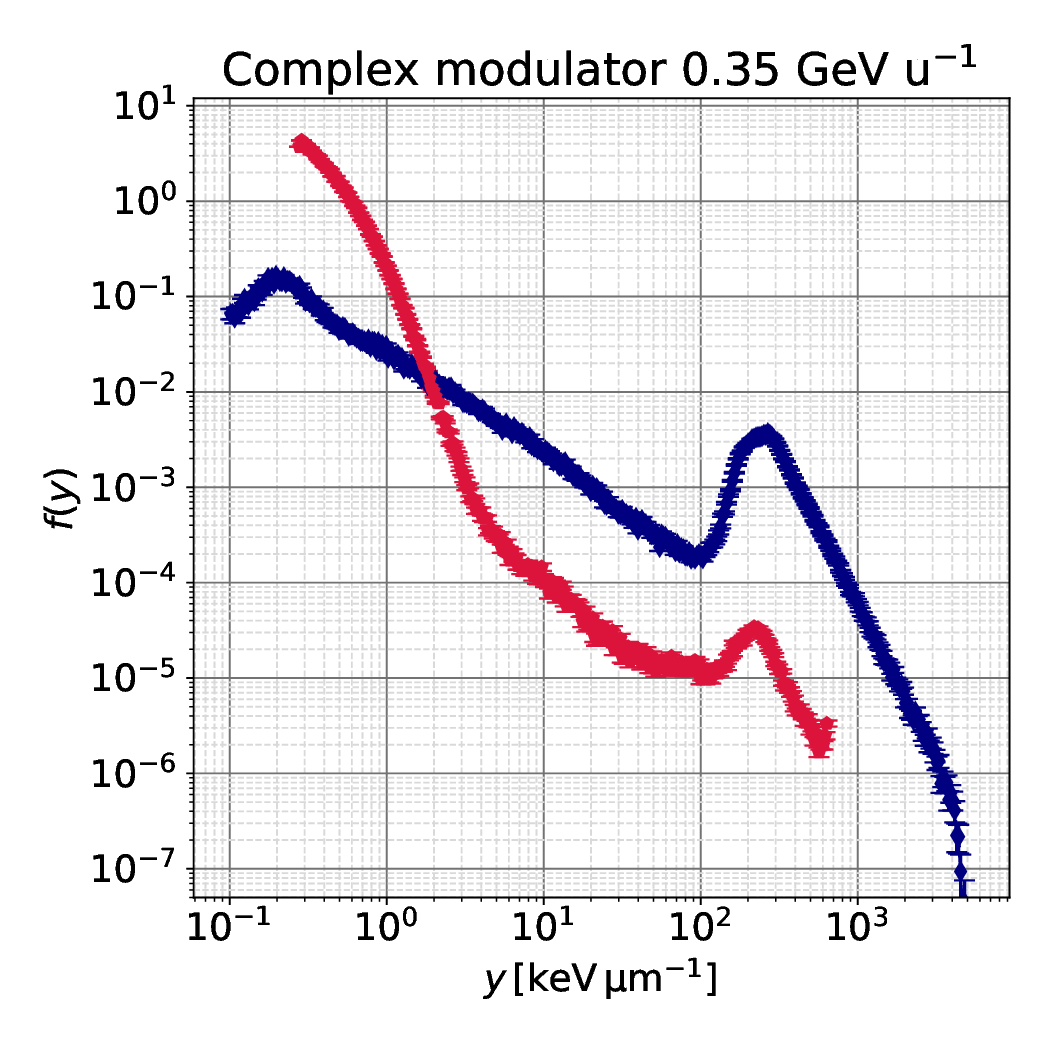}
        \caption{}
        \label{fig:results:0.35GeV_mod}
    \end{subfigure}
    \caption{Comparison of experimental (red) and simulated (blue) $f(y)$ distribution for the three complex modulators configurations. \replyTo{16}Errors are estimated using the statistical error.}
    \label{fig:results:fy_modulators}
\end{figure*}

Microdosimetry was used to characterize each of the six configurations comparing the $f(y)$ microdosimetric spectrum measured experimentally with the one from the MC simulations, as described in \autoref{ssec:Monte Carlo simulations}.
\autoref{fig:results:fy_modulators} shows the comparison for the first three configurations of the GCR simulator, namely, the complex modulators. 
Although the experimental distribution closely follows the simulated one for the \qty{1}{\GeV \per \atomicmassunit} and \qty{0.7}{\GeV \per \atomicmassunit} complex modulators, as shown in \autoref{fig:results:1GeV_mod} and \autoref{fig:results:0.7GeV_mod}, respectively, the agreement between the experimental and simulated distributions for the \qty{0.35}{\GeV \per \atomicmassunit} modulator in \autoref{fig:results:0.35GeV_mod} is not as good.
A defining characteristic common to all the microdosimetric spectra of \autoref{fig:results:fy_modulators} is a broad peak in the lineal energy region roughly delimited by the $100$ - \qty{300}{\keV \per \micro\m} interval. 
The peak is mainly attributable to the energy released by the primary particles. Additionally, it can be observed that, as the energy of the primary beam decreases, the position of the peak shifts towards higher values of lineal energy. This is consistent with theoretical expectations, as lower primary beam energies result in higher energy deposition in the detector.
At lower lineal energy, below the peak region, all the complex modulator microdosimetric spectra feature a steep increase.
This is the result of lighter particles produced by nuclear fragmentation whose contribution in energy deposition is considerably less in the detector with respect to the primary particles.
In this region, a deviation between simulation and experimental data is observed for all complex modulator configurations, with the \qty{0.35}{\GeV \per \atomicmassunit} configuration (\autoref{fig:results:0.35GeV_mod}) showing the largest deviation particularly in the $\lesssim \qty{8}{\keV \per \micro \m}$ lineal energy range.

\begin{figure*}[ht]
    \centering
    \begin{subfigure}[t]{0.32\textwidth}
        \centering
        \includegraphics[width=\textwidth]{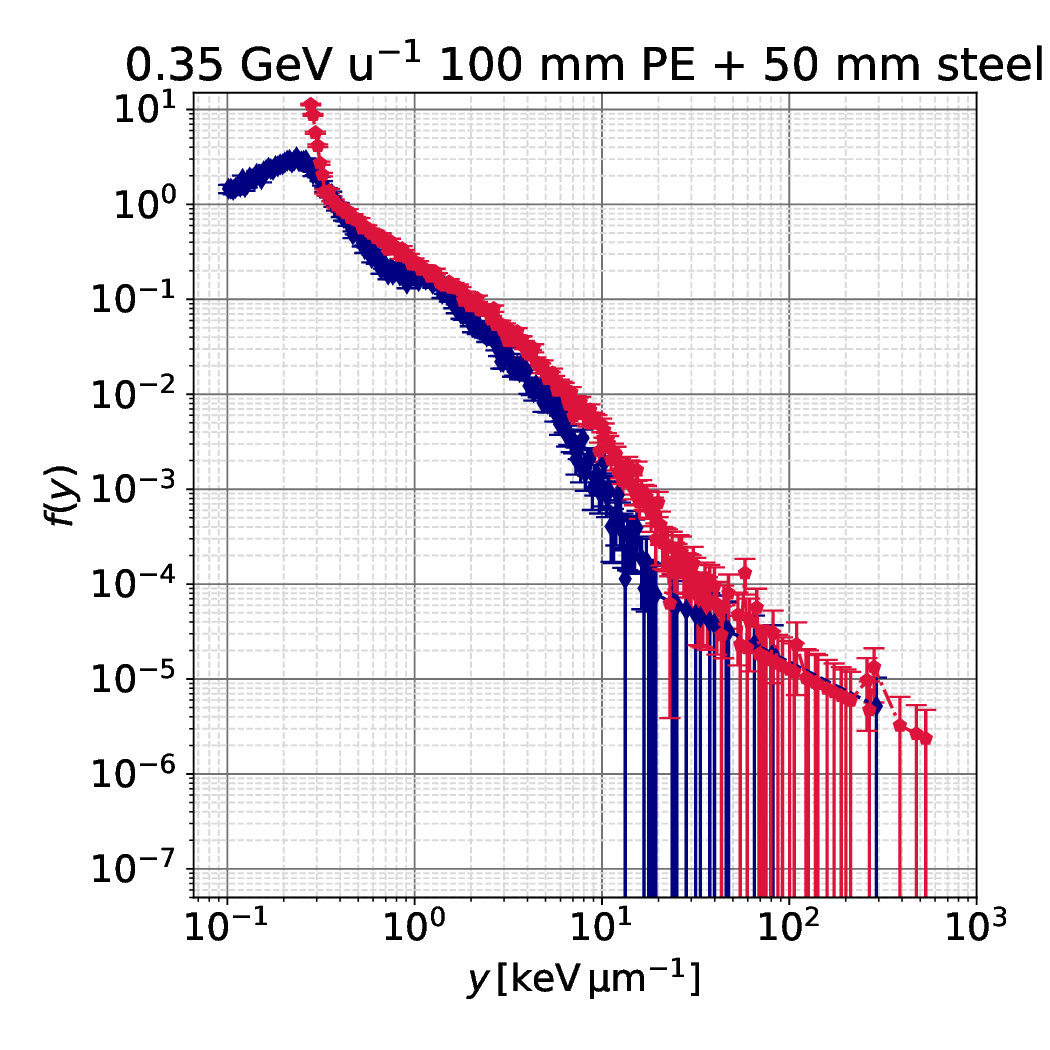}
        \caption{}
    \end{subfigure}
     \hfill
    \begin{subfigure}[t]{0.32\textwidth}
        \centering
        \includegraphics[width=\textwidth]{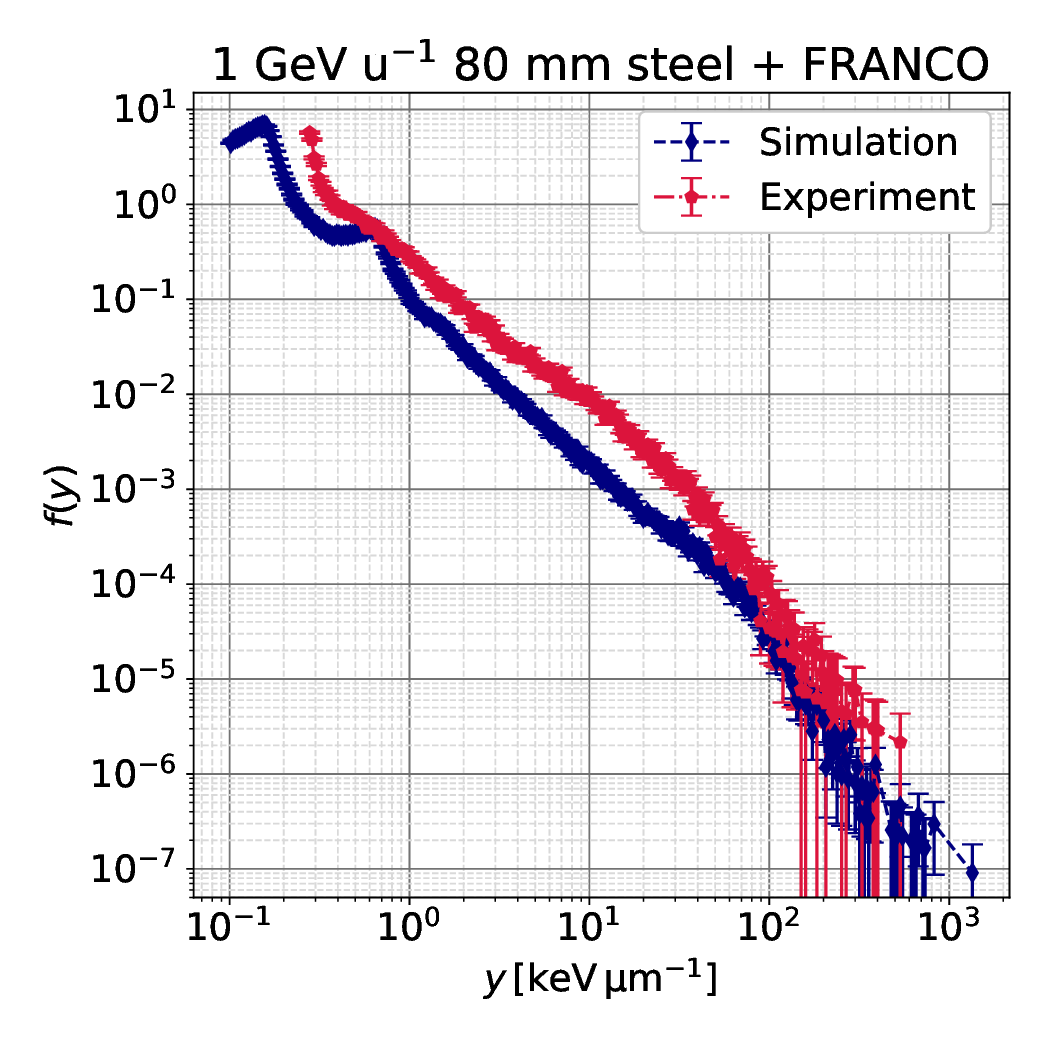}
        \caption{}
    \end{subfigure}
     \hfill
    \begin{subfigure}[t]{0.32\textwidth}
        \centering
        \includegraphics[width=\textwidth]{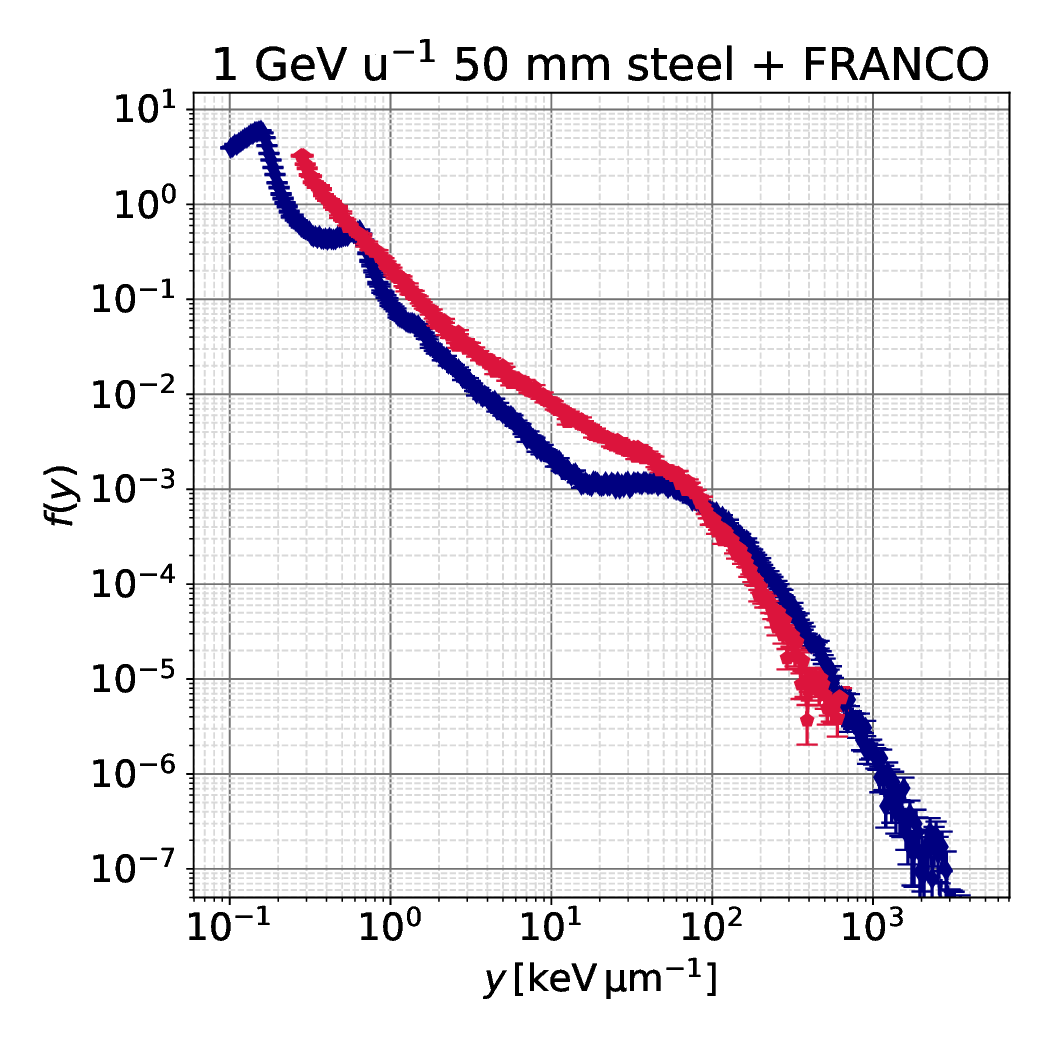}
        \caption{}
    \end{subfigure}
    \caption{Comparison of experimental (red) and simulated (blue) $f(y)$ distribution for the three slab modulator configurations. \replyTo{16}Errors are estimated using the statistical error.}
    \label{fig:results:fy_slab}
\end{figure*}

\autoref{fig:results:fy_slab} shows the remaining three slab modulator configurations. 
A satisfactory agreement between the measurements and simulations can be observed.
Moreover, the three spectra are sharing the same general behavior and, consistently with the design of such configuration that is supposed to fully stop the primary beam in the slab modulators, no peak can be observed contrary to the complex modulator configurations of \autoref{fig:results:fy_modulators}.
On the other hand, it is possible to observe, in a similar manner, a steep increase towards lower values of lineal energy due to the fragments' contributions.

\subsection{Quality factors}
Quality factors were calculated from the $d(y)$ spectra of the six configurations according to the Kellerer-Hahn revision of ICRU 40 and ICRP 60 definitions reported in \autoref{eq:methods:quality_KH} and \autoref{eq:methods:quality_ICRP} respectively. 
Formally, the quality definition $Q_y(y)$ can be applied to lineal energy $d(y)$ distributions, while ICRP 60, $Q_L(L)$, has to be applied to the linear energy transfer (LET)-derived distributions. However, if the lineal energy $y$ is assumed as an approximation of the LET, i.e., $L \simeq y$, the ICRP 60 definition can be used.
As a matter of comparison, both values, $Q_y(y)$ and $Q_L(L = y)$, are calculated for each experimental configuration and compared against values derived from the simulations, as reported in \autoref{tab:results:quality}.
\replyTo{5+6+16}Compatible values of quality factors between simulations and experimental data are obtained in the \qty{0.35}{\GeV \per \atomicmassunit} complex modulator and the slab modulator \qty{1}{\GeV \per \atomicmassunit}  \qty{80}{\mm} steel + FRANCO using the $Q_y(y)$ definition. 
The remaining configurations shows a systematically higher quality factor derived from simulated values with the exception of the \qty{0.35}{\GeV \per \atomicmassunit} slab modulator \qty{50}{\mm} steel + \qty{100}{\mm} PE. \\ 
\replyTo{1}The uncertainties reported in \autoref{tab:results:quality} are derived through propagation of the statistical errors. 
\replyToNew{6}It can be observed that the level of uncertainty associated with the slab modulator configurations is comparatively higher than the one associated with the complex modulator configurations. This effect can be attributed to the combination of a similar number of primary particles being delivered and the lower secondary production yield resulting from nuclear fragmentation specific to the slab target configurations. 
The overall effect reflects the statistical uncertainties, which consequently affects the error estimation of the calculated quality factor.
While statistical uncertainty does not necessarily represent the true magnitude of total error, an alternative formulation combined with the approximation 
$L\simeq y$ allows for a more comprehensive assessment of the error introduced by adopting a different quality factor definition.

\begin{table*}[tb]
\centering
\small
\begin{tabular}{l|ll|ll|}
\multirow{2}{*}{\textbf{Configuration}} &
  \multicolumn{2}{c|}{$Q_y(y)$} &
  \multicolumn{2}{c|}{$Q_L(L = y)$} \\ \cline{2-5} 
 &
  \multicolumn{1}{c|}{\textit{experiment}} &
  \multicolumn{1}{c|}{\textit{simulation}} &
  \multicolumn{1}{c|}{\textit{experiment}} &
  \multicolumn{1}{c|}{\textit{simulation}} \\ \hline
Complex modulator &
  \multicolumn{1}{c|}{\multirow{2}{*}{$25.4 \pm 0.2$}} &
  \multicolumn{1}{c|}{\multirow{2}{*}{$27.4 \pm 0.1$}} &
  \multicolumn{1}{c|}{\multirow{2}{*}{$23.5 \pm 0.2$}} &
  \multicolumn{1}{c|}{\multirow{2}{*}{$25.1 \pm 0.1$}} \\
\qty{1}{\GeV \per \atomicmassunit} &
  \multicolumn{1}{l|}{} &
   &
  \multicolumn{1}{l|}{} &
   \\ \hline
Complex modulator &
  \multicolumn{1}{c|}{\multirow{2}{*}{$26.8 \pm 0.2$}} &
  \multicolumn{1}{c|}{\multirow{2}{*}{$28.4 \pm 0.1$}} &
  \multicolumn{1}{c|}{\multirow{2}{*}{$23.1 \pm 0.2$}} &
  \multicolumn{1}{c|}{\multirow{2}{*}{$24.1 \pm 0.1$}} \\
\qty{0.7}{\GeV \per \atomicmassunit} &
  \multicolumn{1}{l|}{} &
   &
  \multicolumn{1}{l|}{} &
   \\ \hline
Complex modulator &
  \multicolumn{1}{c|}{\multirow{2}{*}{$15.0 \pm 0.1$}} &
  \multicolumn{1}{c|}{\multirow{2}{*}{$14.94 \pm 0.03$}} &
  \multicolumn{1}{c|}{\multirow{2}{*}{$13.9 \pm 0.1$}} &
  \multicolumn{1}{c|}{\multirow{2}{*}{$15.03 \pm 0.03$}} \\
\qty{0.35}{\GeV \per \atomicmassunit} &
  \multicolumn{1}{l|}{} &
   &
  \multicolumn{1}{l|}{} &
   \\ \hline
Slab modulator \qty{0.35}{\GeV \per \atomicmassunit} &
  \multicolumn{1}{c|}{\multirow{2}{*}{\hphantom{0} $4.5 \pm 0.5$}} &
  \multicolumn{1}{c|}{\multirow{2}{*}{\hphantom{0} $1.5 \pm 0.3$}} &
  \multicolumn{1}{c|}{\multirow{2}{*}{\hphantom{0} $4.0 \pm 0.5$}} &
  \multicolumn{1}{c|}{\multirow{2}{*}{\hphantom{0} $1.6 \pm 0.3$}} \\
\qty{50}{\mm} steel + \qty{100}{\mm} PE &
  \multicolumn{1}{l|}{} &
   &
  \multicolumn{1}{l|}{} &
   \\ \hline
Slab modulator \qty{1}{\GeV \per \atomicmassunit} &
  \multicolumn{1}{c|}{\multirow{2}{*}{\hphantom{0} $9.8 \pm 0.4$}} &
  \multicolumn{1}{c|}{\multirow{2}{*}{$10.1 \pm 0.2$}} &
  \multicolumn{1}{c|}{\multirow{2}{*}{\hphantom{0}$8.6 \pm 0.4$}} &
  \multicolumn{1}{c|}{\multirow{2}{*}{\hphantom{0}$9.3 \pm 0.2$}} \\
\qty{80}{\mm} steel + FRANCO &
  \multicolumn{1}{l|}{} &
   &
  \multicolumn{1}{l|}{} &
   \\ \hline
Slab modulator \qty{1}{\GeV \per \atomicmassunit} &
  \multicolumn{1}{c|}{\multirow{2}{*}{$18.0 \pm 0.2$}} &
  \multicolumn{1}{c|}{\multirow{2}{*}{$19.1 \pm 0.1$}} &
  \multicolumn{1}{c|}{\multirow{2}{*}{$16.4 \pm 0.2$}} &
  \multicolumn{1}{c|}{\multirow{2}{*}{$17.9 \pm 0.1$}} \\
\qty{50}{\mm} steel + FRANCO &
  \multicolumn{1}{l|}{} &
   &
  \multicolumn{1}{l|}{} &
   \\ \hline
\end{tabular}
\caption{Quality factor comparison between simulation and experiments for each of the six configurations. Both $Q_y(y)$ according to Kellerer-Hahn revision of the ICRU 40 (\autoref{eq:methods:quality_KH}) and $Q_L(L = y)$ according to ICRP 60 (\autoref{eq:methods:quality_ICRP}) are shown. \replyTo{16}Errors are estimated using the statistical error propagation.}
\label{tab:results:quality}
\end{table*}

\subsection{GCR simulator} \label{sec:results:GCR_simulator}
\begin{figure*}[ht]
    \centering
    \includegraphics[width=\textwidth]{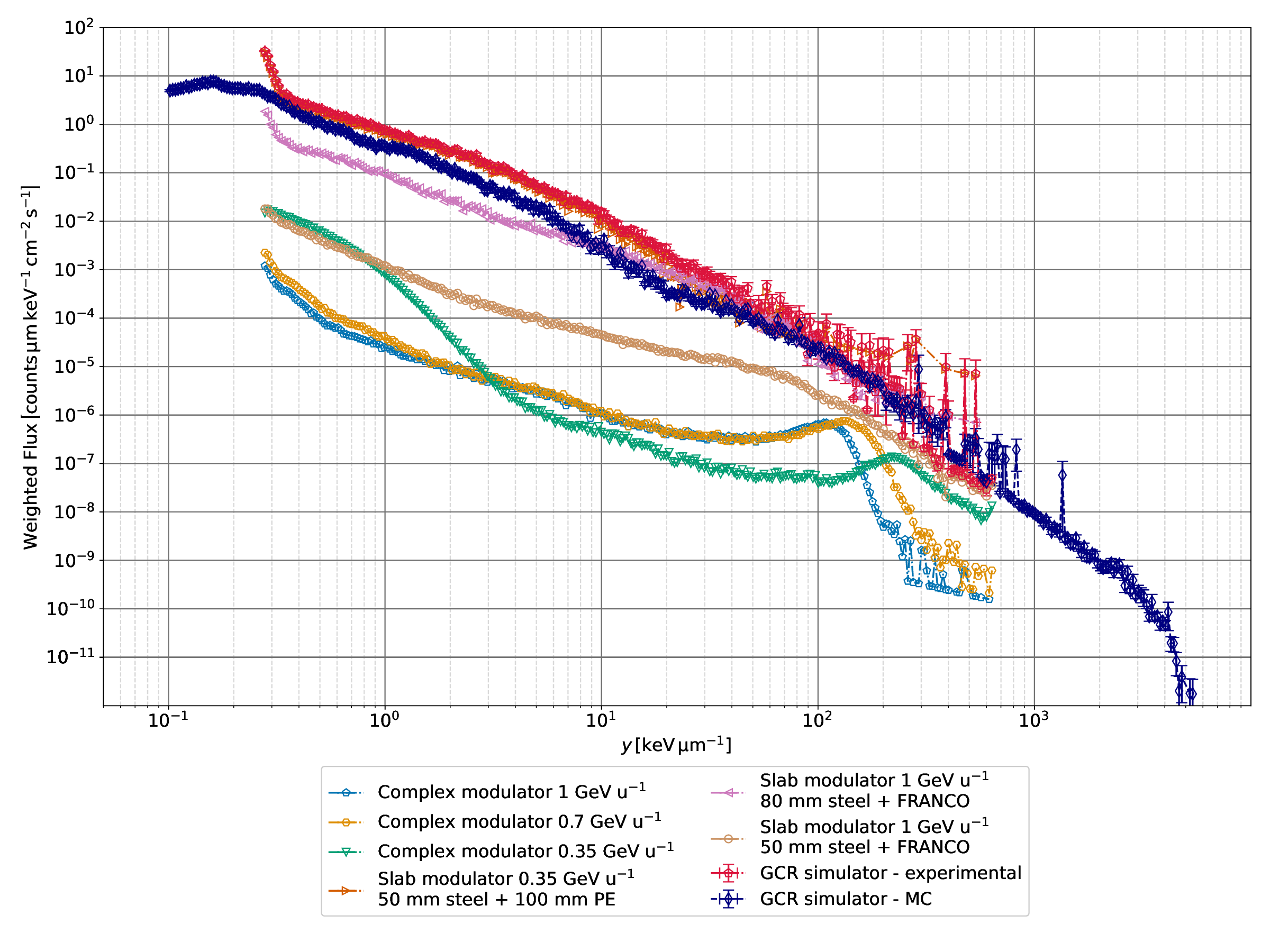}
    \caption{Weighted microdosimetric spectra according to $\omega$ definition in \autoref{eq:results:omega}. The six distinct weighted components, derived from experimental data, are shown separately. The sum of these components is represented by the red line with pentagonal markers. Correspondingly, the blue line with diamond markers illustrates the result obtained from simulated data.  \replyTo{16}Errors are estimated from the statistical error and are shown only for the resulting GCR simulator spectra to avoid overpopulating the plot.}
    \label{fig:results:fy_components}
\end{figure*}

Exposure to the GCR simulator is designed to be delivered by six sequential configurations with an appropriate number of primary particles, as defined by the weights $\omega$: 
\begin{align}
\omega & = \{ \num{9.656e-03}, \, 
    \num{1.322e-02}, \,
    \num{4.718e-03}, \nonumber \\
    &\num{3.621e+04}, \,
    \num{3.944e+02}, \,
    \num{1.002e+00} \} \; \left( \unit{\per \s} \right),
    \label{eq:results:omega}
\end{align}
provided in the same order of \autoref{tab:results:quality}. 
\replyTo{18}These weights are the same of \cite{lunati2025mc}, following the optimization process based on the kinetic energy spectra of different ion species, purely based on physical parameters.
To properly apply the weights following their definition, each lineal energy spectra is normalized by the number of primary particles delivered. 
Additionally, to compare the experimental spectra with the simulated ones, a factor accounting for the different beam field size was also introduced.
This factor was defined as the ratio between the experimentally scanned area by the raster scanning system and the simulation area where the beam is generated.
Finally, before applying the corresponding weights to the spectra, the cross-sectional area of the TEPC detector was used as a common normalization factor for both experimental and simulated spectra.
The result of the six configurations is presented in \autoref{fig:results:fy_components} where the experimental microdosimetric spectra are plotted with their respective weights.
The total spectrum, defined as the sum of the weighted experimental components, is shown in red pentagons. This spectrum represents the radiation field that can be replicated by using the designed sequential irradiation of the GCR simulator.
The same procedure is applied to the simulated spectra, without explicitly showing all the individual components, resulting in the simulated spectrum, in blue diamonds, of \autoref{fig:results:fy_components}.

\subsubsection{Quality factor of the GCR simulator}
The lineal energy spectrum and its associated uncertainty generated by the GCR simulator (\autoref{fig:results:fy_components}), which include both simulated data (blue diamonds) and experimental measurements (red pentagons), enabling the determination of the quality factors and their uncertainties in accordance with \autoref{eq:methods:quality_definition}. These values are reported in \autoref{tab:results:quality_GCR}.
The quality factor directly derived from the LET spectrum after the optimization process, described in \cite{lunati2025mc}, representing the quality factor of the GCR simulator ``by design'', is also reported for comparison in \autoref{tab:results:quality_GCR}.

\begin{table*}[h]
\centering
\small
\begin{tabular}{l|ll|lll|}
 & \multicolumn{2}{c|}{$Q_y(y)$} & \multicolumn{2}{c|}{$Q_L(L = y)$} & \multicolumn{1}{c|}{$Q_L(L)$} \\ \cline{2-6} 
 & \multicolumn{1}{l|}{\textit{experiment}} & \textit{simulation} & \multicolumn{1}{l|}{\textit{experiment}} & \multicolumn{1}{l|}{\textit{simulation}} & \textit{designed}\\ \hline
GCR-simulator & \multicolumn{1}{c|}{$5.9 \pm 0.4$} & \multicolumn{1}{c|}{$5.7 \pm 0.2$} & \multicolumn{1}{c|}{$5.3 \pm 0.4$} & \multicolumn{1}{c|}{$5.3 \pm 0.2$} & \multicolumn{1}{c|}{$5.60 \pm 0.02$}
\end{tabular}
\caption{Quality factor calculated according to the Kellerer-Hahn revision of the ICRU 40 (\autoref{eq:methods:quality_KH}) and ICRP 60 (\autoref{eq:methods:quality_ICRP}). 
\replyToNew{7}The quality factor definitions were evaluated starting from the $f(y)$ (GCR simulator - experimental) and the $f(y)$ simulated (GCR simulator - MC) of \autoref{fig:results:fy_components} obtained by the means of the weights $\omega$ of \autoref{eq:results:omega}.
Additionally, the $Q_L(L)$ is calculated on the LET spectrum directly derived in the optimization process shown in \cite{lunati2025mc} which represents the designed quality factor. \replyTo{16}Errors are estimated using the statistical error propagation.}
\label{tab:results:quality_GCR}
\end{table*}
\replyTo{16}From \autoref{tab:results:quality_GCR} all the calculated values are compatible within the provided error estimation with the exception of the $Q(L=y)$ calculated on simulation data which is not compatible with the designed $Q_L(L)$ obtained from \cite{lunati2025mc}.
The same trend of \autoref{tab:results:quality} is inherited also in \autoref{tab:results:quality_GCR} with the $Q_L(L = y)$ formulation systematically showing smaller central values when compared to the $Q_y(y)$ definition.

\subsubsection{Comparison with microdosimetric data measured in space}
To further validate the GCR simulator, a comparison with a $f(y)$ microdosimetric spectrum measured during the STS-102 Space Shuttle mission has been performed. 
The Shuttle mission was conducted under distinct conditions; therefore, a comparison can be made given the following key differences:  
\begin{itemize}
    \item the detector was placed in the Space Shuttle bay, resulting in different, anisotropic shielding \replyTo{8}between $5-\qty{10} {\g \per \cm \squared}$ of aluminum \cite{Durante2011}.
    \item The mission was performed in a different solar cycle, and, without delving into deep-space\replyTo{8}, however, only the GCR component has been included in the measurement.
    \item The measurement was performed with a different, cylindrical, microdosimeter.
\end{itemize}

\begin{figure*}[htb] 
    \centering
    \includegraphics[width=\textwidth]{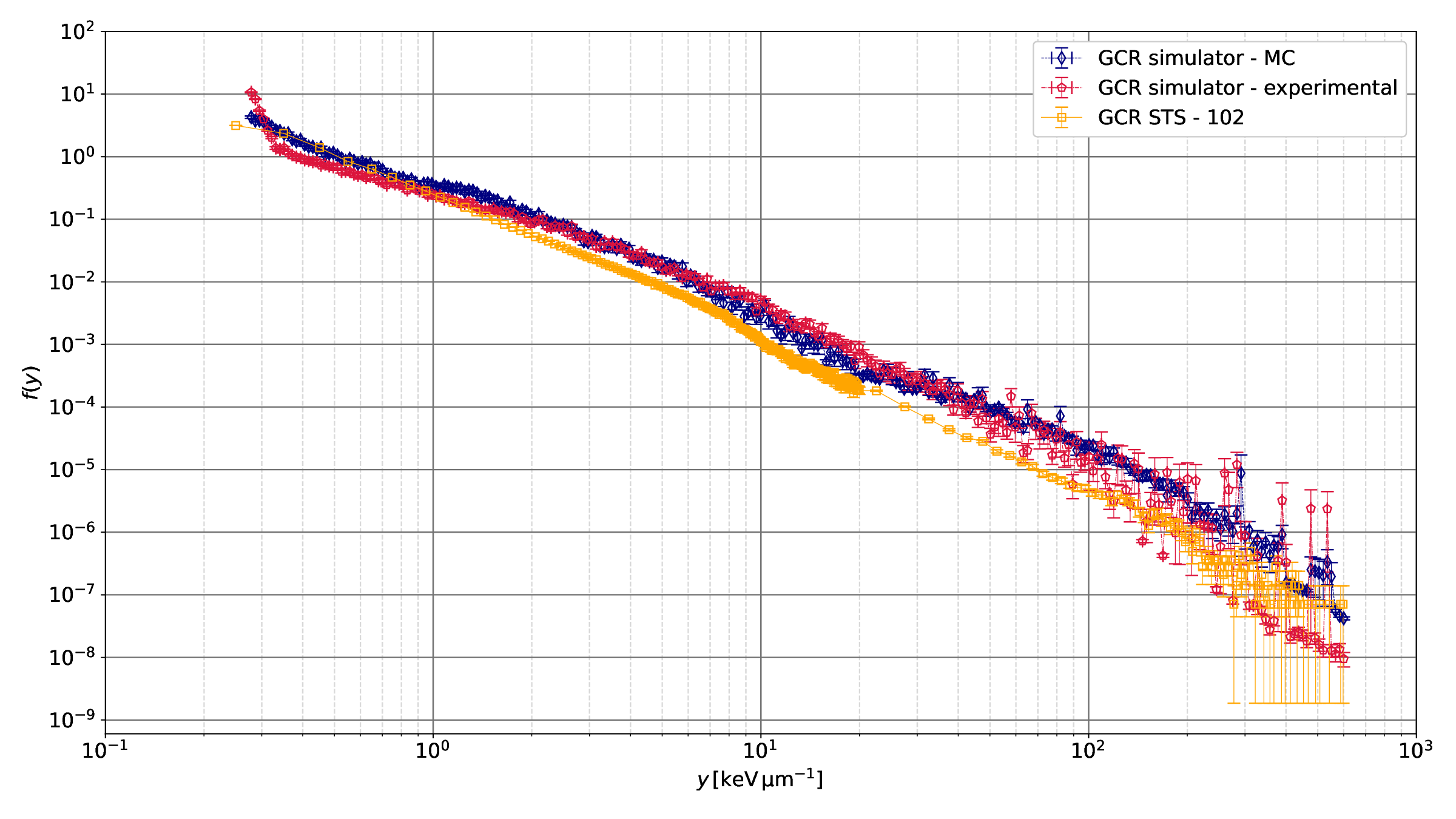}
    \caption{Comparison between GCR simulator experimental (red pentagons) and simulated (blue diamonds) $f(y)$ spectra. Additionally, as a matter of comparison a microdosimetric measurement of GCR acquired with a cylindrical microdosimeter during the STS-102 Space Shuttle mission is shown in gold squares. STS-102 data were obtained from private communication with F. A. Cucinotta.  \replyTo{16}Errors on GCR simulator data have been estimated from the statistical error.}
    \label{fig:results:GCR_vs_STS}
\end{figure*}

To make the measurements compatible, the same lineal energy range was imposed. 
\autoref{fig:results:GCR_vs_STS} shows the experimental data represented by red pentagons, the simulated data by blue diamonds, and the Space Shuttle mission data by gold squares. 
\replyTo{8}It is important to acknowledge that the measurements presented are acquired under different hypothesis. 
However, a reasonable agreement can be observed from \autoref{fig:results:GCR_vs_STS}. The general trend is respected from both the GCR simulator spectra and the STS spectrum. 
In the range with $y < \qty{1}{\keV \per \micro\m}$, the STS is in good agreement with the Monte Carlo data. However, experimental observations exhibit a slight underestimation of this region. 
In the lineal energy range with $\qty{1}{\keV \per \micro\m} \leq y < \qty{100}{\keV \per \micro\m} $, Monte Carlo data and experimental data show a compatible trend systematically above the STS data. 
In the remaining range, the STS measurements are generally compatible within the provided errors to the experimental and simulated GCR simulator data.

\section{Discussion}
To characterize the GCR simulator, microdosimetric measurements with a TEPC were acquired for each of the six individual components.
\replyTo{18}According to the prescribed configurations and weights provided in \cite{lunati2025mc}, it is expected that the particles' abundance and kinetic energy is optimized to match the target GCR simulator. 
Therefore, lineal energy, which is not directly included in the optimization process, is used as an optimization-independent and measurable observable to benchmark the results of the proposed GCR simulator.
By applying the appropriate normalization, accounting for the number of primary particles, scanned area and the weights $\omega$, we could subsequently reconstruct the resulting GCR field, a procedure applicable exclusively to physical experiments, as biological experiments cannot be re-normalized naively.
Consequently, for such experiments the irradiation must be conducted in a well-defined sequence of the configurations, with the correct number of primary particles scaled according to $\omega$ and with the irradiation order as discussed in \cite{lunati2025mc}.\\
The optimized target area spans \qtyproduct{75 x 75}{\mm} \cite{lunati2025mc}; expanding this area would require a larger beam scanning area with wider modulators, which would in turn result in slower irradiation times.
To provide a quantitative estimate, the present GCR simulator implementation, coupled with the SIS-18 synchrotron, can deliver a dose of \qty{300}{\milli \gray}, equivalent to a Mars mission \cite{Simonsen2020}, in less than 30 minutes.
\replyTo{9}This delivery time estimation accounts for beam application of the 6 individual configurations and an additional overhead time required by switching configuration and readying the accelerator parameters to handle a different primary beam. The GSI accelerator can ideally switch energy between extraction cycles (spills), allowing the exchange between configurations to be potentially continuous. 
Moreover, the modulator exchangers' controls employed to implement the experimental setup, in accordance with the GCR simulator configurations, can be incorporated into the accelerator controls software, thereby further automating the configuration exchange. \\
During the delivery of the GCR simulator configurations, it has been demonstrated that wide lineal energy spectra are applied to the target position.
This is a direct consequence of the complex radiation field produced by each GCR simulator configuration, featuring multiple ions with a broad kinetic energy distribution.
In contrast, the NASA approach, utilizes monoenergetic beams of different ions in a sequential manner. Each irradiation made by a single beam resulting with a sharp kinetic energy distribution at the target.
Conversely, by using the presented GCR simulator, the beam application regardless of the configuration, ultimately leads to the instantaneous delivery of a complex radiation field at the target position.
Moreover, the maximum beam energy of \qty{1}{\GeV \per \atomicmassunit}, imposed by the SIS‑18’s technical limits, truncates the simulated GCR kinetic energy spectra, omitting a significant high‑energy component \cite{Slaba2014}.
With the forthcoming facility for antiproton and ion research (FAIR), the maximum deliverable energy will increase to \qty{10}{\GeV \per \atomicmassunit} \cite{Durante2019}, enabling a much more accurate and groundbreaking simulation of the GCR field.

\replyTo{0}When assessing neutron components, the deployed TEPC microdosimeter is not suitable for a thorough characterization of their radiation field, particularly at high energies.
Contrary to the ions species, directly included in the GCR simulator design optimization, neutrons are inevitably produced by the beam interacting with the modulators.
A discussion about the neutron components and their contribution is detailed in \cite{lunati2025mc}. 
The neutron contribution can currently only be quantified using Monte Carlo simulations and their potential impact should be considered when planning future experiments. 
\replyToNew{1}It is therefore the case that a Geant4-based Phase Space is available for public use online (\nolinkurl{https://github.com/chrischu0815/g4_ps_converter}), with the aim of allowing users of the forthcoming GCR simulator to perform simulations as necessary.

By comparing \autoref{fig:results:fy_modulators} and \autoref{fig:results:fy_slab} the widest discrepancy between simulations and experimental results is observable in the \qty{0.35}{\GeV \per \atomicmassunit} complex modulator of \autoref{fig:results:0.35GeV_mod}.
\replyTo{4}Such a configuration represents the most delicate irradiation and setup.
The \qty{0.35}{\GeV \per \atomicmassunit} complex modulator features the lowest energy which results in higher lateral scattering of the primary particles. 
Additionally, the complex modulator is designed to achieve two seemingly opposing objectives: at first, to fully stop the primary beam through carefully engineered geometrical shapes and optimized pin lengths; secondly, to incorporate small holes to preserve the \qty{0.35}{\GeV \per \atomicmassunit} energy component of the primary beam. 
Consequently, the combination of design requirements and low energy of this modulator renders it the most delicate design.
Two possible scenarios can explain the behavior measured in \autoref{fig:results:0.35GeV_mod}: a tilt of the complex modulator making it not perfectly aligned with the beam trajectory and/or residual support material of the printing process in the complex modulator holes.
Although the tilting of the modulator can be investigated directly, modeling the residual paraffin is not as straightforward. In fact, it is expected that the residues will feature a defined spatial pattern near the modulator's hole.
It is hypothesized that combining the effect of 1. and 2. can help to reproduce observed systematic error, as the net effect is the addition of more material along the beam path, therefore limiting the amount of passing primary beam and  increasing nuclear fragmentation, consequentially enhancing the contribution to lower $y$ values in \autoref{fig:results:0.35GeV_mod}.

The quality factor $Q$ obtained from the three complex modulator setups is noticeably higher than the values measured for the \qty{0.35}{\GeV \per \atomicmassunit}  slab modulator and the \qty{1}{\GeV \per \atomicmassunit} \qty{80}{\mm} steel slab modulator configuration. This increase is mainly due to the presence of iron ions, contributing with a higher lineal energy component.
\replyTo{19}Additionally, among the complex modulators, the \qty{0.7}{\GeV \per \atomicmassunit} configuration yields the highest quality factor $Q_y(y)$. This is attributed  to the dumping of the quality function in \autoref{eq:methods:quality_definition} which are penalizing both the very high lineal energies components of the \qty{0.35}{\GeV \per \atomicmassunit} modulator and the comparatively lower lineal energy components of the \qty{1}{\GeV \per \atomicmassunit} modulator.
When considering the slab modulator configurations, $Q$ decreases as more material is placed in the beam path. Such attenuation is caused by nuclear fragmentation of the primary beam, which produces lighter secondary particles.
This observation is consistent regardless of the quality factor definition, $Q_y$ or $Q_L$, and hold true for both simulations and experimentally derived values.
\replyTo{4+19}Notably, when the simulated and experimental data are compared in terms of $Q_y(y)$ for the  \qty{0.35}{\GeV \per \atomicmassunit} complex modulator, the resulting radiation field exhibits compatible quality factors.
Again, this effect can be imputed directly to the definition of the quality functions in \autoref{eq:methods:quality_definition}. These quality functions are designed to significantly dump the low lineal energy contribution region, the same region where discrepancy between simulated and experimental data can be observed from \autoref{fig:results:fy_modulators}. The net effect is that the quality factor value is mostly determined by the higher lineal energy value in correspondence to the peak due to the iron contribution.

As shown in \autoref{tab:results:quality}, simulations underestimate experimental results for the \qty{0.35}{\GeV \per \atomicmassunit} \qty{50}{\mm} steel + \qty{100}{\mm} PE configuration. This setup, with the most material in the beam path, is expected to put a strain on the MC code, particularly in evaluating nuclear cross sections for less common beam and material combinations.
Therefore, the discrepancy can be attributed to the MC transport.

A satisfactory agreement can be found between the simulated and the experimental spectra of the GCR simulator (\autoref{fig:results:fy_components}), resulting from the combination of the six configurations applying the weights $\omega$.
Despite the previously highlighted discrepancy of the \qty{0.35}{\GeV \per \atomicmassunit} complex modulator (\autoref{fig:results:0.35GeV_mod}) between experimental and simulated data, the value of the weight associated to this component, $\qty{4.718e-03}{}$, is sufficiently small to limit its influence on the total GCR spectrum. 
\replyTo{12}This observation is supported by the comparison shown in \autoref{fig:results:fy_components}: as the simulations are not affected by the discrepancy in the low lineal energy range ($y \lesssim \qty{8}{\keV \per \micro \m}$) of \autoref{fig:results:0.35GeV_mod}, a substantial variation would be propagated to the comparison between the GCR simulator experimental spectrum and the simulated one. This is not the case as the general trend of \autoref{fig:results:fy_components} is preserved in both the simulation data and experimental data. Furthermore, the quality factors of \autoref{tab:results:quality_GCR} are compatible between the simulations and the experiment.\\

Finally, when comparing the GCR simulator spectrum with the data collected during the shuttle mission in \autoref{fig:results:GCR_vs_STS},\replyTo{13} 
a general agreement can be observed between the STS and GCR simulator microdosimetric spectra. We emphasize that this is a qualitative comparison under differing conditions. 
While the interpretation of specific sources of discrepancies is challenging, the general agreement provides supporting evidence that the developed GCR simulator produces spectra that are consistent in magnitude and shape with real space-based measurements. 
The total spectrum, obtained using the weights and normalization described in  \autoref{sec:results:GCR_simulator}, are consistent within uncertainties across most of the lineal energy range.

It is worth mentioning that by adjusting the value of the weights $\omega$, as described in \cite{lunati2025mc}, enable to reproduce a GCR field under different heliospheric conditions without changing the hardware.

\replyTo{5+16}Conversely to the previously highlighted discrepancies of the single components of the GCR simulator, between experimental and simulated microdosimetric spectra, \autoref{tab:results:quality_GCR} shows that radiation produced by the GCR simulator is equivalent in terms of quality factor.
This is observed independently from the quality factor definition, $Q_L(L=y)$ or $Q(y)$.
\replyTo{18}When comparing the designed quality factor of \autoref{tab:results:quality_GCR}, all values are compatible with the exception of $Q_L(L=y)$ obtained from simulation data. 
It is worth mentioning that, given the optimization process employed in the design of the GCR simulator, which does not explicitly incorporate biologically derived parameters (e.g., quality factors), the observed agreement is a direct consequence of the physical optimization. Consequentially, this observation confirms the validity of the developed methodology and its validation.
However, attributing the discrepancy observed from the $Q_L(L=y)$ obtained from simulation data to a single, direct cause is not straightforward; nonetheless, it can be stated that the definition of $Q_L$ was applied under the assumption that lineal energy serves as an approximation of the LET measurement.\\
It is important to stress, that lineal energy is a rigorously defined, unambiguous quantity, whereas LET, is a theoretically derived concept whose practical use in experimental contexts is inherently limited. 
To overcome such limitations, and obtain a more comprehensive description of the radiation quality, NASA has developed an alternative approach by defining the quality factor $Q_{\text{NASA}}$, taking into account the knowledge of both effective charge $Z^*$ and velocity $\beta$ of the incident radiation \cite{cucinotta2013space}.
The application of this descriptor is constrained by the practical difficulty of measuring both $Z$ and $\beta$ experimentally. 
In fact, estimation of these quantities necessitates the utilization of advanced telescope detectors, which poses a significant technological challenge, particularly if the system is to be integrated into spacecraft.

\section{Conclusions}

This manuscript presents and validates the performance of the first GCR simulator on European ground. The system employs the sequential irradiation of six different configurations made by ``complex'' and ``slab'' modulators.
By properly scaling the number of primary particles of each setup it is possible to recreate a GCR field in deep space (\qty{1}{\astronomicalunit}), in the 2010 solar minimum conditions, and  after \qty{10}{\g \per \cm \squared} of Al shielding. The capabilities of the system were benchmarked using state-of-the-art MC simulations and experimentally with a TEPC.
\replyTo{14}In our design, six sequential irradiations are used to apply complex radiation fields instantaneously at the target position.
In contrast, the NASA GCR simulator make uses of a sequential irradiation of single monoenergetic ion beams. 
This enables the investigation of the combined effects of different radiation qualities and their interplay on biological systems.
\replyTo{1}At GSI, the Radiobiology Modelling group is advancing a more comprehensive biological characterization. Building on the Local Effect Model (LEM) \cite{Pfuhl2019}, the objective of this study is to broaden the scope of the endpoint outlined in the current manuscript. In addition, a more exhaustive inquiry into the synergistic effects exhibited by the complex radiation fields will be investigated.

For safe and reliable operation, it is of uttermost importance to validate and quality assure the GCR field prior every usage, as a single anomaly potentially can compromise the integrity of all experiments that will be conducted.
Measuring and characterizing the GCR field is a significant challenge directly reflecting its complexity, both in terms of particles populations and broad kinetic energy spectra.
\replyTo{15}In this work, the radiation field was characterized by adopting a microdosimetric approach, a method that has been proven effective over the years at accelerator facilities and numerous space missions.
Nevertheless, providing the best possible description of radiation quality using a compact, easy-to-use detector remains an ongoing issue.
The European Space Agency (ESA) is financially supporting a two-year experimental campaign with the aim of testing and identify multiple detectors systems in such harsh conditions.
Ideally, the outcome of this investigation will identify a device capable of providing not only the best possible characterization of the radiation field, but optimally this device could be integrated into spacecrafts.
Furthermore, measuring all relevant radiation parameters of a complex radiation field is the basis for subsequent biological characterization and modeling.

In conclusion, the GCR simulator will establish a state of the art platform for comprehensive studies of GCR‑induced effects in both electronic circuitry and biological systems. 
The facility is expected to become accessible to experimental groups starting in 2026. 
With these advancements, the GCR simulator will represent an unique platform to perform experiments, helping to bridge the longstanding gap between physical characterization and the resulting biological outcomes.

\section*{Acknowledgements}
The development of the GCR simulator is supported by ESA under contract number 4000102355\allowbreak/10/NL/VJ.
We would also like to acknowledge the additional support by the EU project HEARTS.
HEARTS is a project funded by the European Union under GA No 101082402, through the Space Work Programme of the European Commission.
This research was supported in part by the cluster computing resource provided by the IT Department at the GSI Helmholtzzentrum f{\"{u}}r Schwerionenforschung, Darmstadt, Germany.
The work was performed at the GSI Helmholtzzentrum f{\"{u}}r Schwerionenforschung in Darmstadt (Germany) within the frame of FAIR Phase-0.\\
We would like to thanks Prof. Dr. Francis A. Cucinotta for kindly providing the Space Shuttle data.

\bibliographystyle{elsarticle-num}
\bibliography{references}

\newpage

\appendix
\section{Monte Carlo implementation of the Tissue Equivalent Proportional Counter LET-1/2}
\label{sec:MC_implementation_of_the_TEPC}
\begin{figure}[htb]
    \centering
    \includegraphics[width=0.3\textwidth]{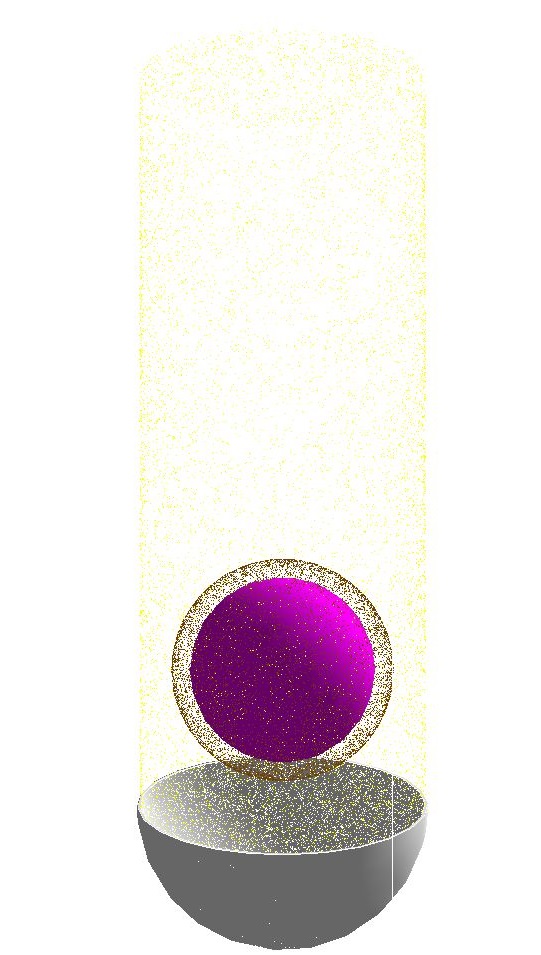}
    \caption{Tissue equivalent proportional counter (TEPC) from Far West Technology model LET-1/2 implemented in Geant4 Monte Carlo toolkit. 
    The yellow dots represents propane gas. The A-150 shell is shown by the brown dots while the active area, filled with propane gas is shown in purple.
    Only the bottom part of the aluminum envelope is shown by the gray semisphere at the bottom of the figure.}
    \label{fig:appendix:TEPC_geometry}
\end{figure}

The Tissue Equivalent Proportional Counter (TEPC) from Far West Technology model LET-1/2 was implemented in Geant4 Monte Carlo toolkit \cite{Agostinelli2003, Allison2006, Allison2016}.
The detector's design is constrained by the geometrical specifications outlined by the constructor, while the material implementation employs the National Institute of Standard and Technology (NIST) database integrated in Geant4.
The detector, show in \autoref{fig:appendix:TEPC_geometry}, is composed of an outer aluminum shell \qty{0.178}{\milli \m} thick. The latter is implemented by combining both a cylinder (omitted for the purpose of illustration) and a semisphere shown in gray color.
Inside the aluminum shell a propane gas with density of $\qty{1.08e-4}{\g \per \cm^3}$ is contained and represented by the yellow dots.
A tissue-equivalent hulled sphere, made by A-150 tissue equivalent material with a thickness of \qty{1.27}{\milli \m} and an inner diameter of \qty{12.7}{\milli \m}, delimited by the brown dots, contains the active volume of the detector. 
Such region, shown in solid purple, is filled by propane gas and collects the energy released by each event of the simulations.
Specific production cuts have been implemented in the detector with the aim of increasing the accuracy of the simulations at the cost of computational time. 
To model the wall effect, a production cut of \qty{1}{\nano \m} was imposed in the A-150 region directly surrounding the sensitive volume. 
Additionally, a maximum step size of $\qty{500}{\nano \m}$ was imposed in the active volume.

\end{document}